\documentstyle[preprint,aps]{revtex}
\begin{document}
\draft
\preprint{ANL-HEP-PR-97-01}
\title{Threshold Resummation of the Total Cross Section \\ 
for Heavy Quark Production in Hadronic Collisions}
\author{Edmond L. Berger and Harry Contopanagos\footnote{ 
Current address: Electrical Engineering Department, University of California  
at Los Angeles, Los Angeles, CA 90024}}
\address{High Energy Physics Division,
             Argonne National Laboratory \\
             Argonne, Illinois 60439, USA \\}
\date{May 26, 1997}
\maketitle

\begin{abstract} 
We discuss calculations of the inclusive total cross section for heavy quark
production at hadron collider energies within the context of perturbative 
quantum chromodynamics, including resummation of the effects of initial-state
soft gluon radiation to all orders in the strong coupling strength.  We resum 
the universal leading-logarithm contributions, and we restrict our 
integrations to the region of phase space that is
demonstrably perturbative.  We include a detailed comparison of the 
differences between ours and other methods.  We provide predictions of the 
physical cross section as a function of the heavy quark mass in 
proton-antiproton reactions at center-of-mass energies of 1.8 and
2.0 TeV, and we discuss estimated uncertainties.
\end{abstract} 
\vspace{0.2in}
\pacs{ }

\section{Introduction and Motivation}
\label{sec:1}

In this report we present and discuss calculations carried out in perturbative 
quantum chromodynamics (QCD) of the inclusive cross section for the production 
of heavy quark-antiquark pairs in hadron reactions\cite
{ref:laeneno,ref:shpaper,ref:lgpaper,ref:catani}.  In most of the paper, we
identify the heavy quark as a top quark, $t$, but the results are valid more 
generally as long at the mass of the quark is sufficiently heavy.  For example, 
they should apply as well to production of a fourth-generation quark, such as 
a postulated $b'$.  

In inclusive hadron interactions at collider energies, 
$h_1 + h_2 \rightarrow t + \bar{t} + X$, $t\bar{t}$ pair production 
proceeds through partonic hard-scattering processes involving initial-state
light quarks $q$ and gluons $g$.  In lowest-order QCD, at 
${\cal O}(\alpha_s^2)$,  the two partonic subprocesses 
are $q + \bar{q} \rightarrow t + \bar{t}$ and $g + g \rightarrow t + \bar{t}$.  
Calculations of the cross section through next-to-leading order, 
${\cal O}(\alpha_s^3)$, involve gluonic radiative corrections to these 
lowest-order subprocesses as well as contributions from the $q + g$ initial 
state\cite{ref:dawson}.  A complete fixed-order calculation at order 
${\cal O}(\alpha_s^n), n \ge 4$ does not exist.  In this paper, we do not 
examine mechanisms for the production of single top quarks or 
antiquarks\cite{ref:single}.

The physical cross section for each production channel is obtained through 
the factorization theorem,
\begin{equation}
\sigma_{ij}(S,m^2)={4m^2\over S}\int_0^{{S\over 4m^2}-1}d\eta\Phi_{ij}\biggl[
{4m^2\over S}(1+\eta),\mu^2\biggr]\hat\sigma_{ij}(\eta,m^2,\mu^2) .
\label{feleven}
\end{equation}
The square of the total hadronic center-of-mass energy is $S$, and the square 
of the partonic center-of-mass energy is $s$.  The mass of the heavy quark 
is $m$, and $\mu$ is the common renormalization/factorization scale of the 
problem.  The variable $\eta={s \over 4m^2} - 1$ measures the distance from the 
partonic threshold.  The indices $ij\in\{q\bar{q},gg\}$ denote the initial 
parton channel.  The partonic cross section 
$\hat\sigma_{ij}(\eta,m^2,\mu^2)$ is obtained commonly from fixed-order QCD
calculations\cite{ref:dawson}, or, as described here, from calculations
that go beyond fixed-order perturbation theory through the inclusion of 
gluon resummation\cite{ref:laeneno,ref:shpaper,ref:lgpaper,ref:catani}.  
The parton flux is 
$\Phi_{ij}(y,\mu^2)=\int_y^1{dx\over x}f_{i/h_1}(x,\mu^2)f_{j/h_2}(y/x,\mu^2)$,
where $f_{i/h_1}(x,\mu^2)$ is the density of partons of type $i$ in hadron 
$h_1$.  We use the notation $\alpha(\mu)\equiv \alpha_s(\mu)/\pi$.  
Unless otherwise specified, $\alpha\equiv \alpha(\mu=m)$ throughout this
paper.  The total physical 
cross section is obtained after incoherent addition of the contributions from 
the the $q\bar{q}$ and $gg$ production channels.  In this paper, we ignore 
the small contribution from the $qg$ channel.

Comparison of the partonic cross section at next-to-leading order with its 
lowest-order value reveals that the ratio becomes very large in the 
near-threshold region.  Indeed, as $\eta \rightarrow 0$, the ``$K$-factor" at 
the partonic level $\hat K(\eta)$ grows in proportion to $\alpha \ln^2(\eta)$. 
An illustration of this behavior may be seen in Fig.~7 of Ref.~[3].  The very 
large mass of the top quark notwithstanding, the large ratio $\hat K(\eta)$ 
makes it evident that the next-to-leading order result does not necessarily 
provide a trustworthy quantitative prediction of the top quark production cross 
section at the energy of the Tevatron collider.  The large ratio casts doubt 
on the reliability of simple fixed-order perturbation theory for physical 
processes for which the near-threshold region in the subenergy variable 
contributes significantly to the physical cross section.  Top quark production 
at the Fermilab Tevatron is one such process, because the top mass is 
relatively large compared to the energy available.  Other examples include 
the production of hadronic jets that carry large values of transverse momentum 
and the production of pairs of supersymmetric particles with large mass.  To 
obtain more dependable theoretical estimates of the cross section in 
perturbative QCD, it is important first to identify and isolate the terms that 
provide the large next-to-leading order enhancement and then to resum these 
effects to all orders in the strong coupling strength.  

We begin in Sec. II with the motivation for the inclusion of the effects of 
intial state soft gluon radiation to all orders in the QCD coupling strength, 
and we review the general formalism of resummation.  In Sec. III, we outline 
the method and domain of applicability of perturbative resummation that we 
developed in the past year\cite{ref:shpaper,ref:lgpaper}.  We present 
predictions in Sec. IV of the physical cross section as a function of the 
heavy quark mass in proton-antiproton reactions at center-of-mass energies of 
1.8 and 2.0 TeV, and we discuss estimated uncertainties.  Our calculation is 
in good agreement with the measured cross section at the reported mass of the 
top quark~\cite{ref:cdfdz}.  At $m$ = 175 GeV and ${\sqrt S}=1.8$ TeV, the 
all-orders resummed cross section is about 9\% greater than the 
next-to-leading order value.  Since the large threshold logarithms are 
mastered by resummation, the theoretical reliability of the resummed result is 
considerably greater than that of the fixed order calculation.  At other values 
of $m$ and ${\sqrt S}$, where the ratio $m/{\sqrt S}$ is larger, the numerical 
effects of resummation can be more significant.  In Secs. V and VI, we compare 
our approach and results with other methods~\cite{ref:laeneno,ref:catani} and 
address criticisms that have been made~\cite{ref:catani}.  The difference 
between our approach and that of Ref.\cite{ref:catani} resides in the 
treatment of subleading logarithmic contributions, and we explain our reasons 
for preferring our method. Conclusions are summarized in Sec. VII.  


\section{Gluon Radiation and Resummation}
\label{sec:2}

The origin of the large threshold enhancement may be traced to initial-state
gluonic radiative corrections to the lowest-order channels.  To avoid 
misunderstanding, we remark that we are calculating the inclusive total cross 
section for the production of a top quark-antiquark pair, i.e., the total 
cross section for $t + \bar{t} + \rm anything$.  The partonic subenergy 
threshold in question is the threshold for $t + \bar{t} + $ any number of 
gluons.  This coincides with the threshold in the invariant mass of the 
$t + \bar{t}$ system for the lowest order subprocesses only.  
 
To specify the kinematic variables, we consider the two-to-three parton 
subprocess $i(k_1)+j(k_2)\rightarrow t(p_1)+{\bar t}(p_2)+g(k)$.  We define the
variable $z$ through the partonic invariants\cite{ref:laeneno}
\begin{equation}
s=(k_1+k_2)^2,\ t_1=(k_2-p_2)^2-m^2,\ u_1=(k_1-p_2)^2-m^2,\ 
(1-z)m^2= s+t_1+u_1.
\label{invariants}
\end{equation}
Alternatively, $(1-z) = {2k \cdot p_1 \over m^2}$.  In the limit that 
$z \rightarrow 1$, the radiated gluon $g(k)$ carries zero momentum.  After 
cancellation of soft singularities and factorization of collinear 
singularities in ${\cal O}(\alpha^3)$, there are left-over integrable 
logarithmic contributions to the cross section associated with initial-state 
gluon radiation.  The contributions of interest here, often expressed in terms 
of ``plus" distributions, are proportional to $\ln(1-z)$.  These logarithmic 
terms are vestiges of the canceled infrared singularities.

The partonic cross section may be expressed generally as
\begin{equation}
\hat{\sigma}_{ij}(\eta,m^2)=\int_{z_{min}}^1
dz{\cal H}_{ij}(z,\alpha)\hat{\sigma}_{ij}'(\eta,m^2,z).
\label{bone}
\end{equation}
We work in the $\overline{\mbox{MS}}$ factorization scheme in which
the $q$, $\bar{q}$ and $g$ densities and the next-to-leading order
partonic cross sections are defined unambiguously.  The lower limit of 
integration, $z_{min} = 1-4(1+\eta)+4\sqrt{1+\eta}$, is set by kinematics.  
The derivative 
$\hat{\sigma}_{ij}'(\eta,m^2,z)=d(\hat{\sigma}_{ij}^{(0)}(\eta,m^2,z))/dz$,
and $\hat{\sigma}_{ij}^{(0)}$ is the lowest-order ${\cal O}(\alpha^2)$ 
partonic cross section expressed in terms of inelastic kinematic variables to 
account for the emitted radiation.

Keeping only the leading logarithmic contributions through 
${\cal O}(\alpha^3)$, we may approximate the total partonic cross section as
\begin{eqnarray}
\hat\sigma_{ij}^{(0+1)}(\eta,m^2)&=&\int_{z_{min}}^1dz
\left\{1+\alpha 2 C_{ij}\ln^2(1-z)\right\}\hat \sigma'_{ij}(\eta,z,m^2)
\nonumber \\
&\equiv& \int_{z_{min}}^1dz{\cal H}^{(0+1)}_{ij}(z,\alpha)\hat\sigma'_{ij}
(\eta,z,m^2)\ ,
\label{twelvep}
\end{eqnarray}
where $C_{q\bar{q}}=C_F=4/3$ and $C_{gg}=C_A=3$. 
As is illustrated in Fig.~1, the leading logarithmic contribution, integrated
over the near-threshold region $1 \ge z \ge 0$, provides an excellent 
approximation to the exact full next-to-leading order physical cross section 
as a function of the heavy quark mass.  

Although a fixed-order ${\cal O}(\alpha^4)$ calculation of 
$t\bar{t}$ pair production does not
exist, we may invoke universality with massive lepton-pair production 
($l\bar{l}$), the Drell-Yan process, to generalize Eq.~(\ref{twelvep}) to 
higher order.  In the near-threshold region, the hard kernel becomes
\begin{equation}
{\cal H}^{(0+1+2)}_{ij}(z,\alpha) \simeq 1+2\alpha C_{ij} \ln^2 (1-z)
+ \alpha^2\biggl[2C^2_{ij} \ln^4 (1-z) - {4\over 3} C_{ij} b_2
\ln^3 (1-z)\biggr] .
\label{btwo}
\end{equation}
The coefficient $b_2=(11C_A-2n_f)/12$, and the number of flavors $n_f=5$. We 
note that the leading logarithmic contributions in each order of perturbation 
theory are all positive in overall sign\cite{ref:sterrev} so that the leading 
logarithm threshold enhancement keeps building in magnitude at each fixed 
order of perturbation theory.  The 
further enhancement of the physical cross section produced by the 
${\cal O}(\alpha^4)$ leading logarithmic terms in the near-threshold region 
is shown in Fig.~1.  At $m =$ 175 GeV, we compute the following ratios of the
physical cross sections in the leading logarithmic approximation:   
$\sigma_{ij}^{(0+1)}/\sigma_{ij}^{(0)} = $1.22, and \linebreak
$\sigma_{ij}^{(0+1+2)}/\sigma_{ij}^{(0+1)} =$ 1.14.  

The goal of gluon resummation is to sum the series in $\alpha^n \ln^{2n}(1-z)$ 
to all orders in $\alpha$ in order to obtain a more trustworthy prediction.
This procedure has been studied extensively for the Drell-Yan 
process\cite{ref:stermano}, and good agreement with data is achieved.  
In essentially all resummation procedures, the 
large logarithmic contributions are exponentiated into a function of the QCD 
running coupling strength, itself evaluated at a variable momentum scale that 
is a measure of the radiated gluon momentum.  For example, in the approach of
Laenen, Smith, and van Neerven (LSvN)\cite{ref:laeneno}, the resummed 
partonic cross section is written as 
\begin{equation}
\hat{\sigma}_{ij}^{R;IRC}(\eta,\mu_o)=\int_{z_{min}}^{1-(\mu_o/m)^3}
dz{\rm e}^{E_{ij}(z,m^2)}\hat{\sigma}_{ij}'(\eta,m^2,z),
\label{bthree}
\end{equation}
where the exponent 
\begin{equation}
E_{ij}(z,m^2) \propto C_{ij}\alpha((1-z)^{2/3}m^2)\ln^2 (1-z).
\label{bfour}
\end{equation}
We note that in Eq.~(\ref{bfour}), the strong coupling strength is evaluated
at the variable momentum scale $(1-z)^{2/3}m^2$.   

Different methods of resummation differ in theoretically and phenomenologically
important respects.   The set of purely leading monomials 
$\alpha^n\ln^{2n}(1-z)$ in $\hat{\sigma}_{ij}$ exponentiates 
directly, with $\alpha$ evaluated at a fixed large scale $\mu = m$, as may 
be appreciated from a glance at Eq.~(\ref{btwo}). This simple result does not 
mean that a theory of resummation is redundant, even if only leading 
logarithms are to be resummed.  Indeed, straightforward replacement of the 
term within the brackets of Eq.~(\ref{twelvep}) with the exponential of 
$\alpha 2 C_{ij}\ln^2(1-z)$ would lead to an exponentially divergent integral 
(and therefore cross section) since the coefficient of the logarithm is 
positive.  The naive approach, therefore, fails from the start, and
more sophisticated resummation approaches must be employed, 
involving scaling and Lorentz-transformation properties of the
classes of terms to be summed. The more sophisticated approaches are not 
free from problems, however. Formally, if not explicitly in some approaches, an 
integral over the radiated gluon momentum $z$ must be done over regions in 
which $z \rightarrow 0$.  Therefore, one significant distinction among methods 
has to do with how the inevitable ``non-perturbative" region is handled in each 
case.  Examination of Eqs.~(\ref{bthree}) and (\ref{bfour}) shows
that an infrared singularity is encountered in the soft-gluon limit
$z \rightarrow 1$:  owing to the logarithmic behavior of $\alpha(q^2)$,
$\alpha(q^2) \propto \ln^{-1} (q^2/\Lambda_{QCD}^2)$,
$\alpha((1-z)^{2/3}m^2) \rightarrow \infty$ as $z \rightarrow 1$.  The infrared 
singularity is a manifestation of non-perturbative physics.  In the
approach of LSvN, this divergence of the integrand at the upper limit of 
integration necessitates introduction of the undetermined infrared  
cutoff (IRC) $\mu_o$ in Eq.~(\ref{bthree}), 
with $\Lambda_{QCD} \leq \mu_o \leq m$.  The cutoff prevents the integration 
over $z$ from reaching the Landau pole of the QCD running coupling constant.  
The presence of an extra scale spoils the renormalization 
group properties of the overall expression.  The unfortunate dependence of the 
resummed cross section on this undetermined cutoff is important numerically 
since it appears in an exponent\cite{ref:laeneno}.  Theoretical uncertainties 
are not easy to evaluate quantitatively in a method that relies on an 
undetermined infrared cutoff.  


\section{Perturbative Resummation}
\label{sec:3}

The method of resummation we employ\cite{ref:shpaper,ref:lgpaper} is
based on a perturbative truncation of principal-value resummation (PVR).  
The principal-value method\cite{ref:stermano}
has an important technical advantage in that it does not require arbitrary
infrared cutoffs, as all Landau-pole singularities are by-passed by a Cauchy
principal-value prescription.  Because extra undetermined scales are absent,
the method also permits an evaluation its perturbative regime of 
applicability, i.e., the region of the gluon radiation phase
space where resummed perturbation theory should be valid.  

To illustrate how infrared cutoffs are avoided in the PVR method, it is useful 
to begin with an expression in moment ($n$) space for the exponent that 
resums the $\ln (1-z)$ terms\cite{ref:stermanold}.  Factorization and 
evolution lead directly to exponentiation in moment space:  
\begin{equation}
E(n, m^2)= -\int\limits^1_0 dx {{x^{n-1}-1}\over{1-x}}
\int\limits^1_{(1-x)^2} {{d\lambda}\over{\lambda}} g
\left[ \alpha\left( \lambda m^2\right) \right].
\label{bsix}
\end{equation}
The function $g(\alpha)$ is calculable perturbatively, but the
behavior of $\alpha(\lambda m^2)$ leads to a divergence of the integrand when
$\lambda m^2 \rightarrow \Lambda_{QCD}^2$.  To tame the divergence, a cutoff
can be introduced in the integral over $x$ or directly in momentum space, in
the fashion of LSvN.  In the principal-value redefinition of resummation, the 
singularity is avoided by replacement of the integral over the real axis $x$ in
Eq.~(\ref{bsix}) by an integral in the complex plane along a contour $P$ that 
has the same endpoints and is symmetric under reflections across the real axis:
\begin{equation}
E^{PV}(n, m^2)\equiv -\int\limits_P
d\zeta {{\zeta^{n-1}-1}\over{1-\zeta}}
\int\limits^1_{(1-\zeta)^2}
{{d\lambda}\over{\lambda}}g\left[\alpha\left(\lambda m^2\right)\right].
\label{bseven}
\end{equation}
The function $E^{PV}(n, m^2)$ is finite since the Landau pole singularity
is by-passed.  Moreover,
$\lim_{n\rightarrow\infty}E^{PV}(n,m^2)=-\infty$, and, therefore,
the corresponding partonic cross section is finite as $z\rightarrow 1
\ (n\rightarrow +\infty)$.
 In Eq.~(\ref{bseven}), all large soft-gluon threshold
contributions are included through the two-loop running of $\alpha$.

Equations~(\ref{bsix}) and~(\ref{bseven}) have identical perturbative 
content, but they have different non-perturbative content since the infrared 
region is treated differently in the two cases.  The non-perturbative content 
is not a prediction of perturbative QCD.  In our study 
of top quark production, we choose to use the exponent only in the 
region of phase space in which the perturbative content dominates.

We use the attractive finiteness of Eq.~(\ref{bseven}) to derive a 
perturbative asymptotic representation of $E(x,\alpha(m))$ that is 
valid in the moment-space interval
\begin{equation}
1<x\equiv \ln n< t\equiv {1\over 2\alpha b_2}.
\label{tseven}
\end{equation}
This perturbative asymptotic representation is
\begin{equation}
E_{ij}(x,\alpha)\simeq E_{ij}(x,\alpha,N(t))=
2C_{ij}\sum_{\rho=1}^{N(t)+1}\alpha^\rho
\sum_{j=0}^{\rho+1}s_{j,\rho}x^j\ .
\label{teight}
\end{equation}
Here
\begin{equation}
s_{j,\rho}=-b_2^{\rho-1}(-1)^{\rho+j}2^\rho c_{\rho+1-j}(\rho-1)!/j!\ ,
\label{tnine}
\end{equation}
and $\Gamma(1+z)=\sum_{k=0}^\infty c_k z^k$, where $\Gamma$ is the Euler gamma 
function.  
The number of perturbative terms $N(t)$ in Eq.~(\ref{teight}) is
obtained\cite{ref:lgpaper} by optimizing the asymptotic approximation
\begin{equation}
\bigg|E(x,\alpha)-E(x,\alpha,N(t))\bigg|={\rm minimum}. 
\label{review1}
\end{equation}
Because of the range of validity in Eq.~(\ref{tseven}) and owing to the 
optimization Eq.~(\ref{review1}), terms in the exponent 
of the form $\alpha^k\ln^kn$ are of order unity, and terms with fewer powers
of logarithms, $\alpha^k\ln^{k-m}n$, are negligible.  The optimization assures 
us that the coefficients of the various terms are benign.  
Resummation is completed in a finite number of steps.  With 
a two-loop expression for the running coupling strength, all monomials of the 
form $\alpha^k\ln^{k+1}n,\ \alpha^k\ln^kn$ are produced in the exponent of 
Eq.~(\ref{teight}). 
Because of the restricted leading-logarithm universality between the 
$t\bar{t}$ and $l\bar{l}$ processes, we discard monomials of the form 
$\alpha^k\ln^kn$ in the exponent.   

The exponent we use is the truncation
\begin{equation}
E_{ij}(x,\alpha,N)=2C_{ij}\sum_{\rho=1}^{N(t)+1}\alpha^\rho s_\rho x^{\rho+1} ,
\label{tseventeen}
\end{equation}
with the coefficients
$s_\rho\equiv s_{\rho+1,\rho}=b_2^{\rho-1}2^\rho/\rho(\rho+1)$.  The number of
perturbative terms $N(t)$ is a function of only the top quark mass $m$.  This
expression contains no factorially-growing (renormalon) terms.  The 
perturbative region of phase space is far removed from the part of phase space
in which renormalons could be influential.  

In Fig.~2 we illustrate the validity of the asymptotic approximation 
for a value of t corresponding to $m=175$ GeV.  Optimization
works perfectly, with $N(t)=6$, 
and the plot demonstrates the typical breakdown of the asymptotic
approximation if $N$ is allowed to increase beyond $N(t)$. This rise  
represents the exponential growth of the infrared (IR) renormalons,
the $(\rho-1)!$ growth in the second term of Eq.~(\ref{tnine}). 
As long as $n$ is in the interval of Eq.~(\ref{tseven}),
all the members of the family in $n$ are optimized 
at the same $N(t)$, showing that the optimum number of 
perturbative terms is a function of $t$, i.e., of $m$ only.

It is valuable to stress that we can derive the perturbative expressions,
Eqs.~(\ref{tseven}), (\ref{teight}), and (\ref{tnine}), from the unregulated
exponent Eq.~(\ref{bsix}) without the PVR prescription, although with less 
certitude.   We discuss this point in some detail in 
Sec. III.B of our long paper\cite{ref:lgpaper}.

After inversion of the Mellin transform from moment space to the physically 
relevant momentum space, the resummed partonic cross sections, 
including all large threshold corrections, can be written in the form of 
Eq.~(\ref{bone}), but with the hard kernel replaced by the resummed form 
\begin{equation} 
{\cal H}^{R}_{ij}(z,\alpha)=\int_0^{\ln({1\over 1-z})}
dx{\rm e}^{E_{ij}(x,\alpha)}
\sum_{j=0}^\infty Q_j(x,\alpha)
\ .\label{bafour}
\end{equation}
The leading large threshold corrections are contained in the exponent 
$E_{ij}(x,\alpha)$, a calculable polynomial in $x$.  The functions 
$\{Q_j(x,\alpha)\}$ arise from the analytical inversion of the Mellin 
transform from moment space to momentum space. 
These functions are produced by the resummation and are expressed in 
terms of successive derivatives of $E$:
$P_k(x,\alpha)\equiv\partial^k E(x,\alpha)/k! \partial^k x$.  Each 
$Q_j$ contains $j$ more powers of $\alpha$ than of $x$ so that 
Eq.~(\ref{bafour}) embodies a natural power-counting of threshold 
logarithms.  

The functional form of $E_{ij}$ for $t\bar{t}$ production is identical to that 
for $l\bar{l}$ production, except for the identification of the two separate 
channels, denoted by the subscript $ij$.  However, only the {\it leading} 
threshold corrections are universal. Final-state gluon radiation as well as
initial-state/final-state interference effects produce subleading logarithmic 
contributions that differ for processes with different final states.  
Accordingly, there is no physical basis for accepting the validity of the 
particular subleading terms that appear in Eq.~(\ref{bafour}).  Among 
all $\{Q_j\}$ in Eq.~(\ref{bafour}), only the very leading one is universal.
This is the linear term in $P_1$ contained in $Q_0$, that turns out to be 
$P_1$ itself.  Since we intend to resum only the universal leading logarithms, 
we retain only $P_1$.  Hence, Eq.~(\ref{bafour}) can be integrated explicitly, 
and the resummed version of Eq.~(\ref{bone}) is
\begin{equation}
\hat{\sigma}_{ij}^{R;pert}(\eta,m^2)=\int_{z_{min}}^{z_{max}}dz
{\rm e}^{E_{ij}(\ln({1\over 1-z}),\alpha)}
\hat{\sigma}_{ij}'(\eta,m^2,z).\label{bthreep}
\end{equation}

We have inserted an upper limit of integration, $z_{max}$, in 
Eq.~(\ref{bthreep}).  This upper limit is set by the boundary
between the perturbative and non-perturbative regimes.  An intuitive
definition of the perturbative region, where inverse power terms are
unimportant, is provided by the inequality 
$\Lambda_{QCD} \over{(1-z)m}$ $\le 1$.
This inequality is identical to the expression in moment space, 
Eq.~(\ref{tseven}),  with the identification $n = {1 \over 1-z}$.  In
momentum space, the same condition is 
realized by the constraint that all $\{Q_j\},\ j\ge 1$ be small 
compared to $Q_0$.  From the explicit expressions\cite{ref:lgpaper} for 
the $\{Q_j\}$, one may show that this constraint corresponds to 
\begin{equation}
P_1\left(\ln\left({1\over 1-z}\right),\alpha\right)\le1\ .
\label{pertmom}
\end{equation} 
Equation~(\ref{pertmom}) is equivalent to the requirement that terms that
are subleading according to perturbative power-counting are indeed subleading
numerically; Eq.~(\ref{pertmom}) is the essence of perturbation theory in this
context.  It assures us that our integration is carried out only 
over a range in which poorly specified subleading terms would not contribute 
significantly even if they had been retained.  

As remarked above, we accept only the perturbative content of
principal-value resummation, and our cross section is evaluated accordingly.  
Specifically, we use Eq.~(\ref{bthreep}) with the upper limit of integration, 
$z_{max}$, calculated from Eq.~(\ref{pertmom}).  The upshot is an effective 
threshold boundary on the integral over the scaled subenergy variable
$\eta$, but one that is calculable, {\it not} arbitrary.  While reminiscent 
perhaps of the cutoff used in the LSvN approach, our threshold boundary has a 
very different and well defined origin.  
Our perturbative resummation probes the threshold down to the point
$\eta\ge \eta_0 =(1-z_{max})/2 $.  Below this value, perturbation theory, 
resummed or otherwise, is not to be trusted.  For a 
top mass $m$ = 175 GeV, we determine that the perturbative regime is
restricted to $\eta \geq$ 0.007 for the $q{\bar q}$ channel and 
$\eta \geq$ 0.05 for the $gg$ channel.  These numbers may be converted to more 
readily understood values of the subenergy above which we 
judge our perturbative approach is valid:  at $m$ = 175 GeV, these are 1.22 GeV 
above the threshold in the $q{\bar q}$ channel, and 8.64 GeV 
above the threshold in the $gg$ channel. The difference reflects the larger
color factor in the $gg$ case.  A larger color factor makes the 
non-perturbative region larger.  (One could attempt to apply 
Eq.~(\ref{bthreep}) all the way to $z_{max} = 1$, i.e., to $\eta =$ 0, but 
one would then be using a {\it model} for non-perturbative
effects, the one suggested by PVR, below the region justified
by perturbation theory.)   We note that the value 1.22 GeV in the $q{\bar q}$ 
channel is comparable to the decay width of the top quark, 
$\Gamma (t \rightarrow b W^+) = 1.55$ GeV.  The width itself provides a 
natural definition of the minimum non-perturbative region.  The two 
independent determinations of the non-perturbative region are in 
agreement\cite{ref:mowidth}.  


\section{Physical Cross Section}
\label{sec:4}

In order to achieve the best accuracy available we wish to include in 
our predictions as much as is known theoretically.  
Our final resummed partonic cross section can therefore be 
written\cite{ref:shpaper,ref:lgpaper}
\begin{equation}
\hat \sigma^{pert}_{ij}(\eta, m^2,\mu^2)=
\hat \sigma^{R;pert}_{ij}(\eta,m^2,\mu^2)-
\hat \sigma^{(0+1)}_{ij}(\eta,m^2,\mu^2)\Bigg|_{R;pert}+
\hat \sigma^{(0+1)}_{ij}(\eta,m^2,\mu^2)\ . 
\label{fthree}
\end{equation}
The second term is the part of the partonic cross section up to one-loop that is
included in the resummation, while the last term is the exact one-loop 
cross section\cite{ref:dawson}.  To obtain physical cross sections, we insert 
Eq.~(\ref{fthree}) into Eq.~(\ref{feleven}), and we integrate over $\eta$.  
Other than the heavy quark mass,
the only undetermined scales are the QCD factorization and renormalization 
scales.  We adopt a common value $\mu$ for both, and we vary this scale over 
the interval $\mu/m\in\{0.5,2\}$ in order to evaluate the theoretical 
uncertainty of the numerical predictions.  We use the CTEQ3M parton 
densities\cite{ref:cteq}.

A quantity of phenomenological interest is the differential cross section 
${d\sigma_{ij}(S,m^2,\eta)\over d\eta}$.  Its integral over $\eta$ is the 
total cross section.  In Fig.~3 we plot these distributions 
for $m=175$ GeV, ${\sqrt S}=1.8$ TeV, and $\mu=m$.   The full range of $\eta$
extends to 25, but we display the behavior only in the near-threshold region
where resummation is important.  We observe that, at the energy of the 
Tevatron, resummation is significant for the $q\bar{q}$ channel and less so 
for the $gg$ channel.  In Fig.~1, the dotted curve shows that our final 
resummed cross section in the $q\bar{q}$ channel, after integration over all
$\eta$, lies about half-way between
the cross sections obtained from the near-threshold leading logarithms at
orders ${\cal O}(\alpha^3)$ and ${\cal O}(\alpha^4)$.  The latter have been 
integrated over the region $0 < z < 1$.    

We display our inclusive total production cross section as a function of the
heavy quark mass in Fig.~4.  The central value of our predictions is defined 
as the value obtained with the 
choice $\mu/m=1$, and the lower and upper limits are  the maximum 
and minimum of the cross section in the range of the hard scale 
$\mu/m\in\{0.5,2\}$.  This definition of the central value is common, but
it results here in an asymmetric uncertainty estimate; the extent of the range 
above the central value is smaller than that below.
At $m =$ 175 GeV, the full width of the uncertainty band 
is about 10\%\ .  In Fig.~5, we show the variation of our resummed cross 
section as the value of the renormalization/factorization scale $\mu$ is 
changed.  As is to be expected, less variation with $\mu$ is evident in the 
resummed cross section than in the next-to-leading order cross section, also 
shown in Fig.~5.  We remark that the cross section reaches its maximum at 
a value of $\mu$ just slightly larger than $m/2$.        
We consider that the variation of the cross section over 
the range $\mu/m\in\{0.5,2\}$ provides a good overall estimate
of uncertainty.  For comparison, we note that over the same range of $\mu$, 
the strong coupling strength $\alpha$ varies by $\pm10$\%\ at $m$ = 175 GeV.  
Using a different choice of parton densities\cite{ref:mrsa}, we find a 4\%\ 
difference in the central value of our prediction\cite{ref:shpaper} at 
$m =$ 175 GeV.  A comparison of the predictions\cite{ref:lgpaper} in the 
$\overline{\mbox{MS}}$ and DIS factorization schemes also shows a modest
difference at the level of $4\%$.

In estimating uncertainties, we do not consider 
explicit variations of our non-perturbative boundary, expressed through 
Eq.~(\ref{pertmom}).  For a fixed $m$ and $\mu$, Eq.~(\ref{pertmom})
is obtained by enforcing dominance of the leading 
hard kernel (determined through perturbative power-counting) over the
subleading ones, all of which are calculable. Therefore, Eq.~(\ref{pertmom})
is {\it derived} and is not a source of uncertainty.  However, at fixed $m$, 
the boundary necessarily varies as $\mu$ and thus $\alpha$ vary.   

Our prediction of the cross section in Fig.~4 is in agreement 
with the data on top quark production\cite{ref:cdfdz}.  We find
$\sigma^{t\bar{t}}(m=175\ {\rm GeV},\sqrt{S}=1.8\ {\rm TeV})=
5.52^{+0.07}_{-0.42}\ pb$.  The central value of this cross section 
is larger than the next-to-leading order value at $\mu=m$ by about $9\%$.  

Extending our calculation at $\sqrt{S}=1.8\ {\rm TeV}$ 
to much larger values of $m$ than shown in Fig.~4,
we find that resummation in the principal $q\bar{q}$ channel produces 
enhancements over the next-to-leading order cross section of $21\%$, $26\%$, 
and $34\%$, respectively, at $m =$ 500, 600, and 700 GeV.  The reason for the
increase of the enhancements with mass at fixed energy is that the threshold 
region becomes increasingly dominant.  Since the $q\bar{q}$ 
channel also dominates in the production of hadronic jets at very large values 
of transverse momenta, we suggest that on the order of $20\%$ of the excess
cross section reported by the CDF collaboration\cite{ref:cdfjets} may well be 
accounted for by resummation.

The top quark cross section increases quickly with the energy of the
$p \bar{p}$ collider.  We provide predictions in Fig.~6 for an upgraded 
Tevatron operating at $\sqrt{S}=2$ TeV.  We determine 
$\sigma^{t\bar{t}}(m=175\ {\rm GeV},\sqrt{S}=2\ {\rm TeV})=
7.56^{+0.10}_{-0.55}\ pb$.  The 2 pb increase in the predicted top quark cross 
section over its value at $\sqrt{S}= $1.8 TeV is about a 37\%\ gain.
The central value rises to 22.4 pb at 
$\sqrt{S}=3\ {\rm TeV}$ and 46 pb at $\sqrt{S}=4\ {\rm TeV}$.  For a fixed mass 
of the heavy quark, the fraction of the cross section supplied 
by the $gg$ subprocess increases rapidly.  For $m = 175$ GeV, this 
fraction is about 15\% at $\sqrt{S}=2$ TeV and 51\% at $\sqrt{S}=4$ TeV.

Turning to $pp$ scattering at the energies of the Large Hadron Collider (LHC)
at CERN, we note a few significant differences from $p\bar{p}$ scattering at 
the energy of the Tevatron.  The dominance of the $q {\bar q}$ production
channel is replaced by $g g$ dominance at the LHC.  Owing to the much larger 
value of $\sqrt{S}$, the near-threshold region in the subenergy variable is 
relatively less important, reducing the significance of initial-state soft
gluon radiation.  Lastly, physics in the region of large subenergy, where 
straightforward next-to-leading order QCD is also inadequate, becomes
significant for $t\bar{t}$ production at LHC energies.  Using the approach
described in this paper, we estimate
$\sigma^{t\bar{t}}(m=175\ {\rm GeV},\sqrt{S}=14\ {\rm TeV})= $ 760 pb.


\section{Other Methods of Resummation}
\label{sec:5}

The groups of Laenen, Smith, and van Neerven (LSvN) and of Catani, Mangano, 
Nason, and Trentadue (CMNT) have also published predictions for the total 
cross section based on resummation of initial state soft gluon radiation.  At
$m=175\ {\rm GeV}$ and $\sqrt{S}=1.8\ {\rm TeV}$, the three values are:
$\sigma^{t\bar t}$({\rm BC}\cite{ref:shpaper,ref:lgpaper}) = 
$5.52^{+0.07}_{-0.42}$ pb;
$\sigma^{t\bar t}$({\rm LSvN}\cite{ref:laeneno}) = $4.95^{+0.70}_{-0.40}$ pb; 
and \linebreak
$\sigma^{t\bar t}$({\rm CMNT}\cite{ref:catani}) = $4.75^{+0.63}_{-0.68}$ pb.  
From the purely numerical point of view, all three predictions agree 
within their estimates of theoretical uncertainty.  However, the resummation 
methods differ, the methods for estimating the 
uncertainties differ, and different parton sets are used.  Comparing with 
LSvN\cite{ref:laeneno},  we find that our central values are 
$10-14\%$ larger, and our estimated theoretical uncertainty is 
$9-10\%$ compared with their $28\%-20\%$.  The larger central value is 
attributable, in part, to the use of different parton densities; our Born 
cross section is about $3-5\%$ larger than the LSvN Born cross section.  
However, it is the choice of the infrared cutoff $\mu_o$ in the LSvN method 
that controls the size of their cross section.  The cutoff $\mu_o$ is selected 
so that the resummed cross section is about equal to the 
next-to-next-to-leading order leading-logarithm cross section 
$\sigma_{ij}^{(0+1+2)}$, obtained from Eq.~(\ref{btwo}).  In contrast, in our 
approach the non-perturbative boundary $z_{max}$ is derived within the context 
of the 
calculation by the requirement that the universal leading-logarithmic terms be 
dominant.  There is no {\it a priori} reason that our resummed result should 
be only 10\% greater than the next-to-leading order cross section at  
$m=175\ {\rm GeV}$ and $\sqrt{S}=1.8\ {\rm TeV}$.  As such, we regard the 
approximate agreement of our result and that of LSvN as somewhat fortuitous.  
Both the central value and the band of uncertainty of the LSvN predictions are 
sensitive to their infrared cutoffs, as we described 
previously\cite{ref:lgpaper}.  

From a theoretical point of view, study of the variation of the predicted 
cross section with the hard scale $\mu$, illustrated here in Fig.~5, 
is important because it reflects the 
stability of the calculation under changes of a perturbative but not directly 
determinable renormalization-factorization scale.  One of the advantages of a 
resummation calculation should be diminished dependence of the cross section on 
$\mu$, less variation than is present in a fixed-order calculation.  To 
estimate theoretical uncertainty, we use the standard $\mu$ variation, and we 
find a band of uncertainty of about 10\% at $m=175\ {\rm GeV}$ and 
$\sqrt{S}=1.8\ {\rm TeV}$.  The LSvN group obtain their uncertainty primarily 
from variations of their infrared cutoff whose role is to measure ignorance 
of non-perturbative effects in that approach.  

The group of Catani, Mangano, Nason, and Trentadue (CMNT)\cite{ref:catani} 
calculate a central value of the resummed cross section (also with 
$\mu/m = 1$) that is less than $1\%$ above the exact next-to-leading order 
value.  There are similarities and differences between our approach to 
resummation and the method of Ref.~\cite{ref:catani}.  We both begin in moment 
space with the same universal 
leading-logarithm expression, but differences occur after the transformation to
momentum space.  In this paper, we set aside comments on mathematical aspects 
of their procedure and focus instead on phenomenological issues of interest.  
As remarked above, the Mellin transformation generates subleading terms in 
momentum space.  The suppression of the effects of resummation arises from 
the retention in Ref.~\cite{ref:catani} of numerically significant
non-universal subleading logarithmic terms. 

CMNT choose to retain all of these inasmuch as they perform 
the Mellin inversion numerically.  Instead, in keeping with the fact that
subleading logarithmic terms are not universal, we retain only the universal
leading logarithm terms in momentum space, and we restrict our phase
space integration to the region in which the subleading terms would not be 
numerically significant regardless.  The differences in the two approaches can 
be stated more explicitly if we examine the perturbative expansion of the
kernel ${\cal H}^{R}_{ij}(z,\alpha)$, Eq.~(\ref{bafour}).  If, instead of 
restricting the resummation to the universal leading logarithms only, we were 
to use the full content of Eq.~(\ref{bafour}), we would arrive at an 
analytic expression that is equivalent to the numerical inversion of 
Ref.~\cite{ref:catani}, 
\begin{equation}
{\cal H}^{R}_{ij} \simeq 1+2\alpha C_{ij} 
\biggl[\ln^2 (1-z) + 2\gamma_E \ln (1-z)\biggr] + {\cal O}(\alpha^2) ;
\label{padovao}
\end{equation} 
where $\gamma_E$ is Euler's constant.  In terms of this expansion, in our work 
we retain only the leading term $\ln^2 (1-z)$ at order 
$\alpha$, but CMNT retain both this term and the subleading term 
$ 2\gamma_E \ln (1-z)$.  Indeed, if the subleading 
term $ 2\gamma_E \ln (1-z)$ is discarded in Eq.~(\ref{padovao}), the
residuals $\delta_{ij}/\sigma_{ij}^{NLO}$ defined in Ref.~\cite{ref:catani} 
increase from 
$0.18\%$ to $1.3\%$ in the $q\bar{q}$ production channel and from $5.4\%$ to 
$20.2\%$ in the $gg$ channel\cite{ref:MLMPN}.  After addition of the two 
channels, the total residual $\delta/\sigma^{NLO}$ grows from the negligible 
value of about $0.8\%$ cited in Ref.~\cite{ref:catani} to the value 
$3.5\%$.  While still smaller than 
the increase of about $9\%$ that we obtain, the increase of $3.5\%$ vs. $0.8\%$ 
shows the substantial influence of the subleading logarithmic terms retained
in Ref.~\cite{ref:catani}. 

We judge that it is preferable to integrate over only the region of phase 
space in which the subleading term is suppressed numerically.  Our reasons 
include the fact that the subleading term is not universal, is not the same as 
the subleading term in the exact ${\cal O}(\alpha^3)$ calculation, and can be 
changed if one elects to keep non-leading terms in moment space.  The 
subleading term is 
negative and numerically very significant when it is integrated throughout 
phase space (i.e., into the region of $z$ above our $z_{max})$.  
In the $q\bar{q}$ channel at $m=175$ GeV and ${\sqrt S}=1.8$ TeV, 
its inclusion eliminates more than half of the contribution from the leading 
term.  In our view, the presence of numerically significant subleading 
contributions begs the question of consistency.  A further justification for 
the retention of only the leading term is that it approximates the exact 
next-to-leading order result well, as shown in Fig.~1.  The choice made in 
Ref.~\cite{ref:catani} reproduces only one-third of the exact next-to-leading 
order result.  The influence of subleading terms is amplified at higher 
orders where additional subleading structures occur in the approach of 
Ref.~\cite{ref:catani} with significant numerical coefficients proportional 
to $\pi^2$, $\zeta(3)$, and so forth. We present a more detailed discussion 
of this issue in the next section.


\section{Further Discussion of the CMNT Approach}
\label{sec:6}

In this section we offer a more systematic analysis of the role played 
in the approach of Ref.~\cite{ref:catani} by non-universal subleading 
logarithms.  We are interested in 
expansions of the resummed momentum-space kernel, Eq.~(\ref{bafour}), up to 
two loops. Therefore, the corresponding
cross sections are integrable down to threshold, $z_{max} = 1$ and $\eta = 0$. 
As we will see, though, the effects of the various classes of logarithms are 
pronounced if one continues the region of integration outside our 
perturbative regime. 

In moment space, the exponent to two-loops is obtained from Eq.~(\ref{teight}):
\begin{equation}
E_{ij}^{[2]}(x,\alpha)=g \alpha(s_{2,1}x^2+s_{1,1}x+s_{0,1})
+g \alpha^2(s_{3,2}x^3+s_{2,2}x^2+s_{1,2}x+s_{0,2}) ,
\label{extraone}
\end{equation}
with $g=2C_{ij}$ and $x = \ln n$.
The corresponding hard kernels in momentum space can be derived from 
Eq.~(\ref{bafour}), according to the formulas (91) through (94) of 
Ref.\cite{ref:lgpaper}.  In the notation of Ref.\cite{ref:lgpaper}, we 
retain terms up to those that are linear in $P_2$.
Alternatively, one can perform the analytical Mellin inversion directly,  
beginning with Eq.~(\ref{extraone}).  The two methods provide identical 
results down to the monomials $x_z^2\alpha^2$; $x_z\equiv \ln(1/(1-z))$.  
Here we quote results based on the explicit inversion of Eq.~(\ref{extraone}). 
(The monomial $x_z\alpha^2$ can be obtained also in the first approach if we 
keep the quadratic term in $P_2$.)
After a trivial integration is performed, the results for the one- and 
two-loop hard kernels are 
\begin{equation}
{\cal H}^{(1)}=x_z^2\alpha\{g s_{2,1}\}
+x_z\alpha\{g(s_{1,1}+2c_1s_{2,1})\}\ ,
\label{extratwo}
\end{equation}
and
\begin{eqnarray} 
& &{\cal H}^{(2)}=x_z^4\alpha^2\{g^2s_{2,1}^2/2\}
+x_z^3\alpha^2\{gs_{3,2}+g^2(s_{2,1}s_{1,1}+
2c_1s_{2,1}^2)\}\nonumber \\
& &+x_z^2\alpha^2\{g(s_{2,2}+3c_1s_{3,2})+g^2(s_{1,1}^2/2+3c_1s_{1,1}s_{2,1}+
s_{2,1}s_{0,1}+  s_{2,1}^2[6c_2-\pi^2]\}\nonumber \\
& &+x_z\alpha^2\{g(s_{1,2}+2c_1s_{2,2}+s_{3,2}[6c_2-\pi^2])\nonumber \\
& &+g^2(s_{0,1}s_{1,1}
+2c_1s_{0,1}s_{2,1}+c_1s_{1,1}^2+s_{2,1}s_{1,1}[6c_2-\pi^2]+s_{2,1}^2[12c_3
-2\pi^2c_1])\}\ .
\label{extrathree}
\end{eqnarray}
All the constants are defined in Eqs.~(\ref{teight}) and (\ref{tnine}).  We 
remark that Eq.~(\ref{extratwo}) includes a leading logarithmic term, 
$x_z^2\alpha$, as well as a next-to-leading term, $x_z\alpha$.  

The question we now address is whether it is justified and meaningful to retain 
all of the terms in Eqs.~(\ref{extratwo}) and (\ref{extrathree}) in the 
computation of the resummed cross section.  The issue has to do with what one 
intends by resummation of leading logarithms.
We use the term {\it leading logarithm} resummation to denote the case in 
which the moment space exponent, Eq.~(\ref{extraone}), contains only the 
constants $E_{LL}=\{s_{\rho+1,\rho},0\}$.  This is also what is done in the 
method of Ref.~\cite{ref:catani}, and the exponent {\it in moment space} in 
their work is identical to that used for our predictions, 
Eq.~(\ref{tseventeen}).  However, in contrast to our expression in momentum 
space, Eq.~(\ref{bthreep}), the corresponding expression in momentum space of 
Ref.~\cite{ref:catani} includes the numerical equivalent of all 
terms in Eqs.~(\ref{extratwo}) and (\ref{extrathree}) that are proportional to 
$s_{\rho+1,\rho}$.

If expressed analytically, the corresponding ``LL" hard kernels in the method 
of Ref.~\cite{ref:catani} are
\begin{equation}
{\cal H}^{(1)}_{LL}=x_z^2\alpha g-x_z\alpha2g\gamma_{E} ,
\label{extrafour}
\end{equation}
and
\begin{eqnarray}
& &{\cal H}^{(2)}_{LL}=x_z^4\alpha^2g^2/2+x_z^3\alpha^2\{2gb_2/3-2\gamma_Eg^2\}
\nonumber \\
& &+x_z^2\alpha^2\{-2gb_2\gamma_E+g^2[3\gamma_E^2-\pi^2/2]\}\nonumber \\
& &+x_z\alpha^2\{2gb_2[3\gamma_E^2-\pi^2/2]/3+g^2[\gamma_E\pi^2-2\gamma^3-
4\zeta(3)]\}\ ,
\label{extrafive}
\end{eqnarray} 
where $\zeta(s)$ is the Riemann zeta function; $\zeta(3)=1.2020569$.
Evaluating the expressions numerically for the $q{\bar q}$ channel, we obtain
\begin{equation}
{\cal H}^{(1)}_{LL}=x_z^2\alpha\times 2.66666-x_z\alpha\times 3.07848 ,
\label{extrasix}
\end{equation}
and
\begin{eqnarray}
{\cal H}^{(2)}_{LL}=x_z^4\alpha^2\times 3.55555
-x_z^3\alpha^2\times 4.80189-x_z^2\alpha^2\times 33.88456-x_z\alpha^2\times
9.82479\ .
\label{extraseven}
\end{eqnarray}
Apart from the leading monomials that are the same as those in our approach, 
Eqs.~(\ref{extrasix}) and (\ref{extraseven}) include a series of 
subleading terms, each of which has a significant negative coefficient. 
In practice, these subleading terms in the approach of 
Ref.~\cite{ref:catani} suppress the effects of resummation 
essentially completely.  One of the effects of this suppression is that the 
resummed partonic cross section is {\it smaller} than its next-to-leading 
order counterpart in the neighborhood of $\eta =$ 0.1, a region in which the 
next-to-leading order partonic cross section takes on its largest values.  This 
point is illustrated in Fig.~3 of the second paper in Ref.~\cite{ref:catani}.  

Although the specific set of subleading terms in 
Eqs.~(\ref{extrasix}) and (\ref{extraseven}) is generated in the inversion of 
the Mellin transform, we would argue that the terms are accidental, at best.  
Our reasoning is based on an examination of the exact next-to-leading order 
calculation of the cross section for heavy quark production and of similar 
calculations of the Drell-Yan process up to two-loops.  
First, terms involving $\gamma_E$ do not appear in the exact next-to-leading 
order calculation of the hard part, since they are removed in the specification 
of the ${\overline{\rm MS}}$ factorization scheme.  Therefore, the term 
proportional to $\gamma_E$ in Eq.~(\ref{extrafour}) is suspect.  Second, if we 
extract the specific value of the subleading logarithm from the full 
${\cal O}(\alpha^3)$ next-to-leading order calculation\cite{ref:dawson}, we 
find~\cite{ref:deriv} $x_z\alpha(2g - 41/6)$ instead 
of the term $-x_z\alpha2g\gamma_{E}$ in 
the equivalent CMNT Eq.~(\ref{extrafour}).  Referring to Eq.~(\ref{extrasix}), 
we remark that instead of the numerical coefficient 3.07848, one would have 
the smaller value 1.5 if the subleading logarithm of the exact 
${\cal O}(\alpha^3)$ calculation were used.  Thus, not only is the 
${\cal O}(\alpha)$ subleading term retained in the approach 
of Ref.~\cite{ref:catani} different from that of the exact calculation, it is 
numerically about twice as large.  Third, we would claim that the results of a 
LL resummation should not rely on the subleading structures in any significant 
way.  However, in the approach of Ref.~\cite{ref:catani}, 
Eq.~(\ref{extrafour}), which is the one-loop projection 
of their resummed prediction, reproduces only 1/3 of the exact 
${\cal O}(\alpha^3)$ enhancement, the other 2/3 being cancelled by the 
second (non-universal) term of Eq.~(\ref{extrafour}).  Correspondingly, the 
method of Ref.~\cite{ref:catani} fails an important consistency check: it 
sets out to resum the 
threshold corrections responsible for the large enhancement of the cross 
section at next-to-leading order; in the end, it does not reproduce most 
of this enhancement.

Addressing questions associated with the $\gamma_E$ terms~\cite{ref:hcvisit}
CMNT examine a type of NLL resummation in the second paper of 
Ref.~\cite{ref:catani}.  In this NNL resummation, the 
$\{s_{\rho+1,\rho},s_{\rho,\rho}\}$ terms are retained in the 
exponent of Eq.~(\ref{extraone}).  The corresponding hard kernels become 
\begin{equation}
{\cal H}^{(1)}_{NLL}=x_z^2\alpha g ,
\label{extraeight}
\end{equation}
and
\begin{equation}
{\cal H}^{(2)}_{NLL}=x_z^4\alpha^2g^2/2+x_z^3\alpha^22gb_2/3
-x_z^2\alpha^2g^2[\gamma_E^2+\pi^2/2]-x_z\alpha^2\{gb_2[2\gamma_E^2+\pi^2/3]
+g^24\zeta(3)\}\ .
\label{extranine}
\end{equation}
Comparing Eqs.~(\ref{extraeight}) and (\ref{extranine}) with 
Eq.~(\ref{btwo}), we observe that Eq.~(\ref{extraeight}) 
is identical to the one-loop projection of our hard kernel.  As shown in 
Ref.\cite{ref:lgpaper}, it yields an excellent approximation to the 
exact next-to-leading 
order cross section.  On the other hand, our two-loop projection contains only 
the first two terms of Eq.~(\ref{extranine}).   The term proportional to 
$x_z^3\alpha^2$ is present in our case, along with the leading term 
proportional to $x_z^4\alpha^2$, because it comes from the leading 
logarithms in the exponent $E(n)$, through one-loop running of the coupling 
strength.  In contrast to 
Eq.~(\ref{extrafive}), Eq.~(\ref{extranine}) relegates the influence 
of the ambiguous constant coefficients to lower powers of $x_z$ (but with 
larger negative coefficients).  In the amended scheme of 
Ref.~\cite{ref:catani}, the unphysical 
$\gamma_E$ terms are still present in the two-loop result, 
Eq.~(\ref{extranine}), along with $\pi^2$ and $\zeta(3)$ terms that may be 
expected but whose coefficients have no well defined physical origin.  
Recast in numerical form, Eqs.~(\ref{extraeight}) and (\ref{extranine}) become  
\begin{equation}
{\cal H}^{(1)}_{NLL}=x_z^2\alpha\times 2.66666 ,
\label{extraninep}
\end{equation}
and
\begin{equation}
{\cal H}^{(2)}_{NLL}=x_z^4\alpha^2\times 3.55555+x_z^3\alpha^2\times 3.40739
-x_z^2\alpha^2\times 37.46119-x_z\alpha^2\times 54.41253\ .
\label{extraten}
\end{equation}
We call attention to the significant difference between the coefficients of 
all but the very leading power of $x_z$ in 
Eqs.~(\ref{extrasix}) and (\ref{extraseven}) with respect to those in  
Eqs.~(\ref{extraninep}) and (\ref{extraten}), and to the fact that the 
numerical coefficients grow in magnitude as the power of $x_z$ decreases.  

Using their NLL amendment, CMNT find that the central value of their resummed 
cross section exceeds the next-to-leading order 
result by $3.5\%$ (both $q{\bar q}$ and $gg$ channels added). This increase 
is about 4 times larger than the central value of the increase obtained in 
their first method, closer to our increase of about $9\%$.   The reason for the 
significant change of the increase resides with the subleading structures, 
viz., in the differences between the LL version 
Eqs.~(\ref{extrasix}) and (\ref{extraseven}) and the NLL version 
Eqs.~(\ref{extraninep}) and (\ref{extraten}).  The subleading terms at 
two-loops cause a total suppression of the two-loop contribution (in fact, 
that contribution is negative), if one integrates all the way into what we 
call the non-perturbative regime.  This suppression explains why an 
enhancement of only $3.5\%$ is obtained in the amended method of 
Ref.~\cite{ref:catani}, rather than our $9\%$.

CMNT argue that retention of their subleading terms in momentum space is 
important for ``energy conservation".  By this statement, they mean that one 
begins the formulation of resummation with an expression in momentum space  
containing a delta function representing conservation of the fractional 
partonic momenta.  In moment space, this delta function subsequently 
unconvolves the resummation.   Therefore, when one inverts the Mellin transform
to return to momentum space, the full set of logarithms generated by this 
inversion are required by the original energy conservation.  This line of 
reasoning would be compelling {\it if the complete exponent $E(n)$ in moment 
space were known exactly}, i.e., if the resummation in moment space were exact 
in representing the cross section to all orders.  However, the exponent is 
truncated in all approaches, and knowledge of the logarithms it resums 
reliably is limited both in moment and in momentum space.  Hence, the set of 
logarithms produced by the Mellin inversion in momentum space should also be 
restricted.  In our approach energy conservation is obeyed in momentum space 
consistently with the class of logarithms resummed.  On the other hand, in the 
method of Ref.~\cite{ref:catani}, knowledge is claimed of all logarithms 
generated from the Mellin inversion, despite the fact that the truncation in 
moment space makes energy conservation a constraint restricted to the class of 
logarithms that is resummable, i.e., a constraint restricted by the truncation 
of the exponent $E(n)$.  The two approaches would be equivalent provided a 
constraint be in place on the effects of subleading logarithms.  This  
constraint is precisely our Eq.~(\ref{pertmom}).  By contrast, no such 
constraint is furnished in Ref.\cite{ref:catani}.  For this reason the results 
of Ref.~\cite{ref:catani} are numerically unstable if one set of the 
logarithms generated in momentum space is adopted as ``the set corresponding 
to energy conservation", and then compared with another set, produced by a 
different truncation of $E(n)$. 

We have identified the terms responsible for the difference between our answer 
for the resummed cross section and that of Ref.~\cite{ref:catani}.   These 
differences reside 
with subleading logarithms whose presence is not substantiated by physical 
arguments.  The essence of our determination of the perturbative regime in 
Eq.~(\ref{pertmom}) is precisely that, in this regime, 
subleading structures are also {\it numerically subleading}, whether or not 
the classes of subleading logarithms coming from different truncation of 
the master formula for the resummed hard kernel, Eq.~(\ref{bafour}), are 
included.  The results presented in Fig.~11 of Ref.\cite{ref:lgpaper}, show 
that if we alter our resummed hard kernel to account for subleading
structures but still stay within our perturbative regime, the resulting 
cross section is reduced by about $4\%$, within our band of perturbative 
uncertainty.

A criticism~\cite{ref:catani} is that of putative ``spurious factorial
growth'' of our resummed cross section, above and beyond the infrared 
renormalons that are eliminated from our approach.  The issue, as we 
demonstrated in 
Eq.~(29) of Ref.\cite{ref:lgpaper}, can be addressed most easily if we 
substitute any monomial appearing in Eq.~(\ref{extrathree}), symbolically 
$\alpha^mc(l,m)\ln^l x_z$, into Eq.~(\ref{bone}) and integrate over $z$:  
\begin{equation}
\alpha^mc(l,m)\int_{z_{min}}^1dz\ln^l x_z=\alpha^mc(l,m)(1-z_{min})l!
\sum_{j=0}^l\ln^j(1/(1-z_{min}))\ .
\label{uclaone}
\end{equation}
For the purposes of this demonstration we set $\hat \sigma'_{ij}=1$.  
The coefficients $c(l,m)$ can be read directly from 
Eq.~(\ref{extrathree}).  For the leading logarithmic terms, 
\begin{equation}
c(2m,m) \propto 1/m! ,
\label{uclaonep}
\end{equation} 
where this factorial comes directly from exponentiation.   After the 
integration over the entire $z$-range, the power of the logarithm in $x_z$ 
becomes a factorial multiplicative factor, $l!$. The presence of $l!$ 
follows directly from the existence of the powers of $\ln x_z$ 
that are present explicitly in the finite-order result in pQCD and is therefore 
inevitable.
If this exercise is 
repeated, but with the range of integration in Eq.~(\ref{uclaone}) constrained 
to our perturbative regime, one obtains the difference between the 
right-hand-side of Eq.~(\ref{uclaone}) and a similar expression containing 
$z_{max}$.  The result is numerically smaller, but both of the pieces are 
multiplied by $l!$.  

The factorial coefficient 
$l!$ is neither the only nor the most important source of enhancement.  
For the leading logarithms at two-loop order, $l=2m=4$, and the overall 
combinatorial coefficient from Eqs.~(\ref{uclaone}) and (\ref{uclaonep}) is 
$(2m)!/m!=12$.  For comparison, at representative values of $\eta$ near 
threshold, $\eta=0.1$ and 0.01, the sum of logarithmic terms in 
Eq.~(\ref{uclaone}) provides factors of 16.1 and 314.3, respectively.  
Similarly, the (multiplicative) color factors at this order of perturbation 
theory are $(2C_{ij})^2=$ 7.1 and 36 for the $q\bar{q}$ and $gg$ channels, 
respectively.  All of these features are connected to the way 
threshold logarithmic contributions appear in finite-order pQCD and how they 
signal the presence of the non-perturbative regime.  Thus, 
preoccupation with the $l!$ factor seems misplaced.

The phrase ``spurious factorial growth'' appears to rename 
the logarithmic enhancements present in Eqs.~(\ref{extratwo}) and 
(\ref{extrathree}), after the integral over $z$.  On the other hand, 
according to our understanding, the claim\cite{ref:catani} 
of ``absence of factorial growth" is based on the use in 
Ref.~\cite{ref:catani} of Eq.~(\ref{extraseven}) for their main predictions,
an expression that contains non-universal subleading logarithms, all with
significant negative coefficients.  Mathematically, factorial growth is present 
for each of the powers of the logarithm in Eq.~(\ref{uclaone}), since 
these monomials are irreducible (linearly independent).  Absence of 
factorial growth based on a numerical cancellation between various classes 
of logarithms, most of them with physically unsubstantiated coefficients, 
appears to us to be an incorrect use of terminology, rather than a transparent
expression of the mathematics.  

From a purely phenomenological point of, one cannot claim 
that a $9\%$ increase of the top quark cross section at $m=175\ {\rm GeV}$ 
and $\sqrt{S}=1.8\ {\rm TeV}$ reveals factorial growth but that an  
$0.8\%$ increase does not.  In the approach taken in Ref.~\cite{ref:catani}, 
the effects of resummation are suppressed by a series of subleading logarithms 
with large negative coefficients.  If there is no physical basis for 
preference of Eqs.~(\ref{extrafour}) and (\ref{extrafive}) 
over Eqs.~(\ref{extraeight}) and (\ref{extranine}), as the authors of 
Ref.~\cite{ref:catani} seem to suggest, then the difference in the 
resulting cross sections can be interpreted as a measure of theoretical 
uncertainty.  This interpretation would not justify firm 
conclusions of a minimal $0.8\%$ increment based on the choice of 
Eqs.~(\ref{extrafour}) and (\ref{extrafive}).   

As remarked in Sec.~IV, the value quoted in Ref.~\cite{ref:catani} for the 
physical cross 
section at $m=175\ {\rm GeV}$ and $\sqrt{S}=1.8\ {\rm TeV}$, including 
theoretical uncertainty, lies within our uncertainty band.  Therefore, the 
numerical differences between us for the specific case of top quark production 
at the Tevatron have little practical significance.  However, there are 
important differences of principal in our treatment of subleading 
contributions that will have more significant consequences for predictions in 
other processes or at other values of top mass and/or at other energies, 
particularly in reactions dominated by $gg$ subprocesses.  


\section{Discussion and Conclusions}
\label{sec:7}

In this paper, we present and discuss the calculation of the inclusive cross 
section for top quark production in perturbative QCD, including the 
resummation of initial-state gluon radiation to all orders in the strong 
coupling strength.  The advantages of the perturbative resummation 
method\cite{ref:shpaper,ref:lgpaper} we espouse are that 
there are no arbitrary infrared cutoffs and there is a well-defined 
perturbative region of applicability where subleading logarithmic terms are
numerically suppressed.  Our theoretical analysis shows that perturbative 
resummation without a model for non-perturbative behavior is both  possible and 
advantageous.  In perturbative resummation, the perturbative region of phase
space is separated cleanly from the region of non-perturbative behavior.

When evaluated for top quark production at ${\sqrt S}=1.8$ TeV,  
our resummed cross sections are about $9\%$ above the next-to-leading order
cross sections computed with the same parton distributions. The 
renormalization/factorization scale 
dependence of our cross section is fairly flat, resulting in a $9-10\%$ 
theoretical uncertainty.  This variation is smaller than the corresponding 
dependence of the next-to-leading cross section, as should be expected.
Our perturbative boundary of 1.22 GeV above the threshold in the dominant 
$q\bar q$ channel is comparable to the hadronic width of the top quark, a 
natural definition of the perturbative boundary.  Neither this, nor the 
somewhat larger value of 8.64 GeV above threshold in the $g g$ channel, 
associated with the larger color factor in the $g g$ channel, is ``unphysically 
large"\cite{ref:catani}.  In recent papers\cite{ref:catani}, 
the authors state that the increase in cross section they find with 
their resummation method is no more than $1\%$ over next-to-leading order.  
The numerical difference in the two approaches 
boils down to the treatment of the subleading logarithms, which can 
easily shift the results by a few percent, if proper care is not
taken. Our approach includes the universal leading
logarithms only while theirs includes non-universal subleading
structures which produce the suppression they find. In Sec. VI, we explain 
why we judge that that our treatment of the subleading structures is 
preferable. 

Our theoretical analysis and the stability of our cross sections under $\mu$
variation provide confidence that our perturbative resummation procedure 
yields an accurate calculation of the inclusive top quark cross section at 
Tevatron energies and exhausts present understanding of the perturbative 
content of the theory.  Our resummed top quark cross section is about $9\%$ 
above the next-to-leading order cross section with an estimated theoretical 
uncertainty of $9-10\%$, associated with $\mu$ variation.  An entirely 
different procedure to estimate the overall theoretical uncertainty is to 
compare our enhancement of the cross section above the next-to-leading order 
value to that of Ref.~\cite{ref:catani}, again yielding about $10\%$.  An 
interesting question is 
whether theory can aspire to an accuracy of better than $10\%$ for the 
calculation of the top quark cross section.  To this end, a more complete 
mastery of subleading logarithms would be desirable, perhaps 
requiring a formidable complete calculation at next-to-next-to-leading order of 
heavy quark production, to establish the possible pattern of subleading 
logarithms, and resummation of both leading and subleading 
logarithms\cite{ref:kidster}.  

Our prediction agrees with data, within the large 
experimental uncertainties.  Despite the different treatment of subleading 
terms, our calculation of the inclusive cross section for top quark production 
at the Fermilab Tevatron and that of Ref.~\cite{ref:catani} fall within the 
estimated uncertainties of each other.  If a cross section significantly 
different from ours is measured in future experiments at the Tevatron with 
greater statistical precision, we would look for explanations in effects  
beyond QCD perturbation theory.  These explanations might include unexpectedly 
substantial non-perturbative effects or new production mechanisms.  An 
examination of the distribution in $\eta$ might be revealing.

In this paper, we concentrate on the all-orders summation of large logarithmic 
terms that are 
important in the near-threshold region of small values of the scaled partonic 
subenergy, $\eta \rightarrow 0$.  Our specific case is top quark production at 
the Fermilab Tevatron collider.  Other processes for which threshold 
resummation and our methodology will also be pertinent include
the production of hadronic jets that carry large values of transverse momentum 
and the production of pairs of supersymmetric particles with large mass.  
There is a complementary region of large $\eta, \eta \rightarrow 1$, in which 
the resummation of different large logarithms may also be important.  The 
production of heavy quarks $Q$ in the limit that the hadronic center-of-mass 
energy is much larger than the quark mass provides an example.  The dominant 
production channel is $gg \rightarrow Q \bar{Q} X$; the ratio of the 
next-to-leading order partonic cross section divided by its leading-order 
approximation is very large at large $\eta$.  Correspondingly, the fixed-order 
cross section will not offer a reliable prediction, and an all-orders approach 
is called for~\cite{ref:collins}.  Particular cases include the total cross 
sections for bottom quark production at the Tevatron and top quark production 
at the CERN Large Hadron Collider.  


\section*{Acknowledgments}

Work in the High 
Energy Physics Division at Argonne National Laboratory is supported by 
the U.S. Department of Energy, Division of High Energy Physics, 
Contract W-31-109-ENG-38.  ELB is grateful to J. Schlereth for valuable 
and timely assistance.   



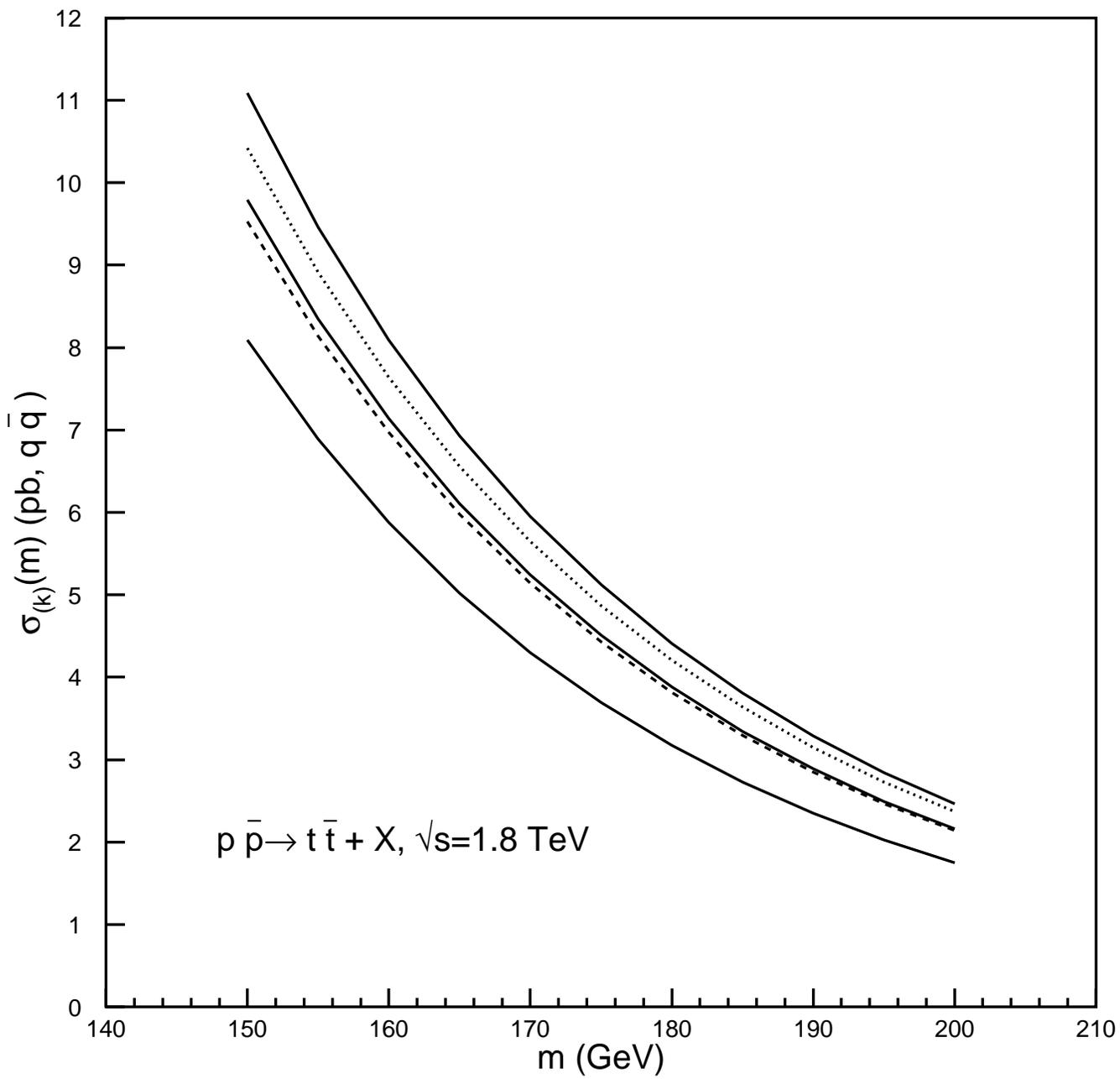
\begin{figure}
\caption{Physical cross sections in the $q\bar{q}$ channel as a 
function of the heavy quark mass, in the $\overline{\rm MS}$ scheme.
The solid lines denote the finite-order partial sums of the
universal leading-logarithmic contributions from the explicit
${\cal O}(\alpha^3)$ and ${\cal O}(\alpha^4)$ calculations for the
$t\bar{t}$ and Drell-Yan processes, respectively.
Lower solid: $\sigma^{(0)}$;
middle solid: $\sigma^{(0+1)}$;
upper solid: $\sigma^{(0+1+2)}$. The dashed curve represents the exact
next-to-leading order calculation for $t\bar{t}$ production,
in excellent agreement with $\sigma^{(0+1)}$. The dotted curve is our
resummed prediction.}
\label{fig1}
\end{figure}

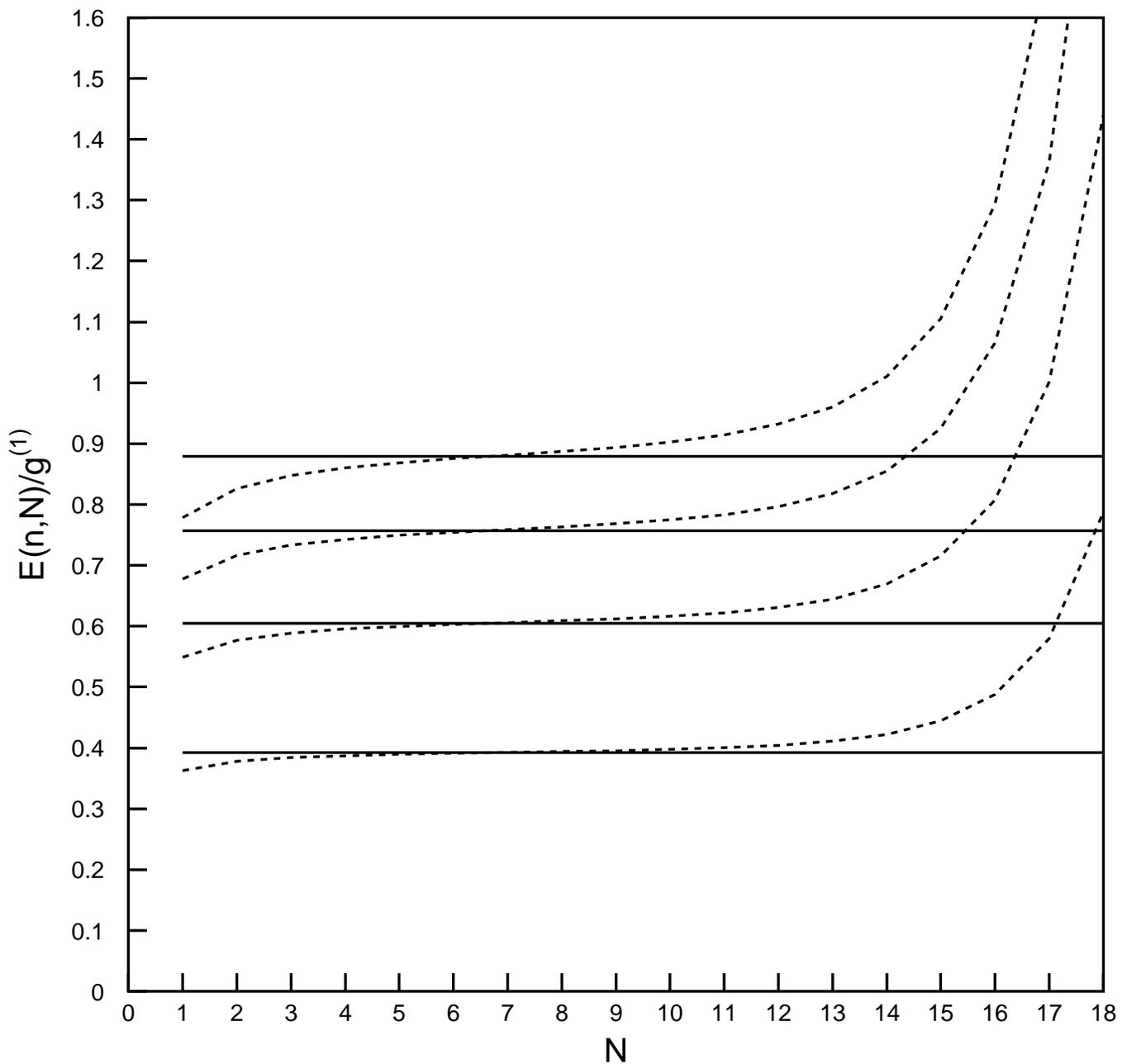
\begin{figure}
\caption{Optimum number of perturbative terms in the exponent
with PVR. The solid family is for PVR and the dashed set for the perturbative
approximation, both families increasing, for parametric values
$n=10,20,30,40$.}
\label{fig2}
\end{figure}

\begin{figure}
\caption{Differential cross sections $d\sigma/d\eta$ 
for $p \bar{p} \rightarrow t \bar{t} X$ at $\protect\sqrt{S}=1.8$ TeV and 
$m =$175 GeV in the $\overline{\rm MS}$-scheme for (a) the $q\bar{q}$ and 
(b) the $gg$  channel:
Born (dotted),  next-to-leading order (dashed), and resummed (solid).}
\label{fig3}
\end{figure}

\begin{figure}
\caption{Inclusive cross section for heavy quark production 
at $\protect\sqrt{S}=$ 1.8 TeV in the
$\overline{\rm MS}$ scheme. The dashed curves show our perturbative
uncertainty band, while the solid curve is our central prediction.} 
\label{fig4}
\end{figure}

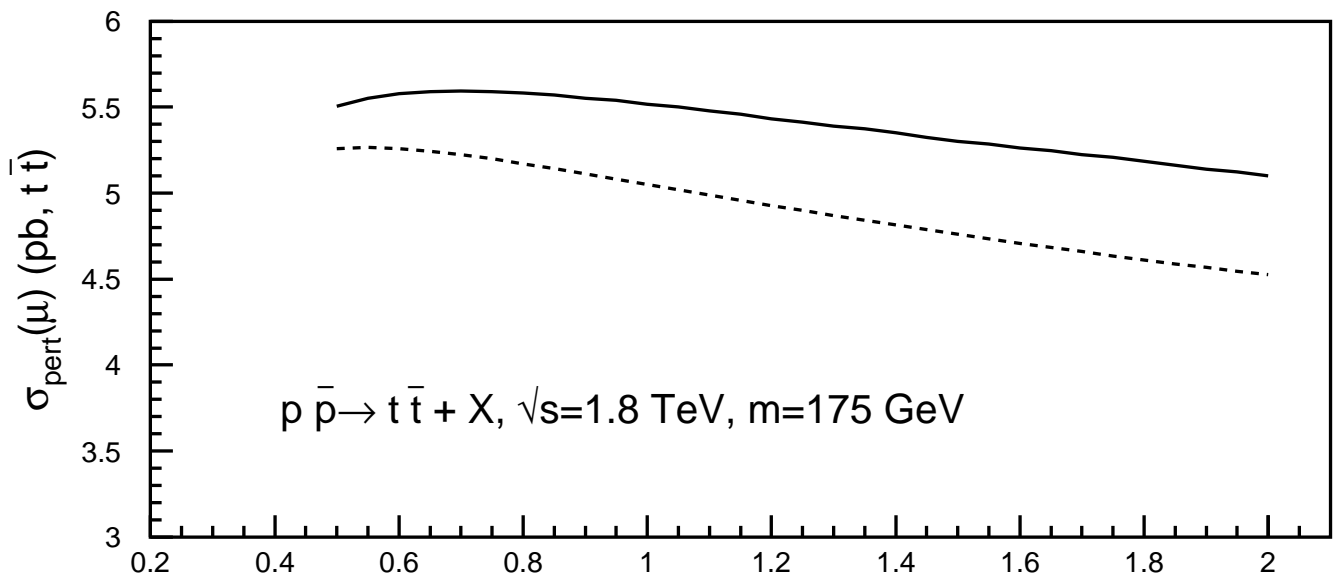
\begin{figure}
\caption{Renormalization/factorization hard scale dependence of the resummed
(solid) and next-to-leading order (dashed) cross sections at 
$\protect\sqrt{S}=$ 1.8 TeV for $m =$175 GeV.}
\label{fig5}
\end{figure}

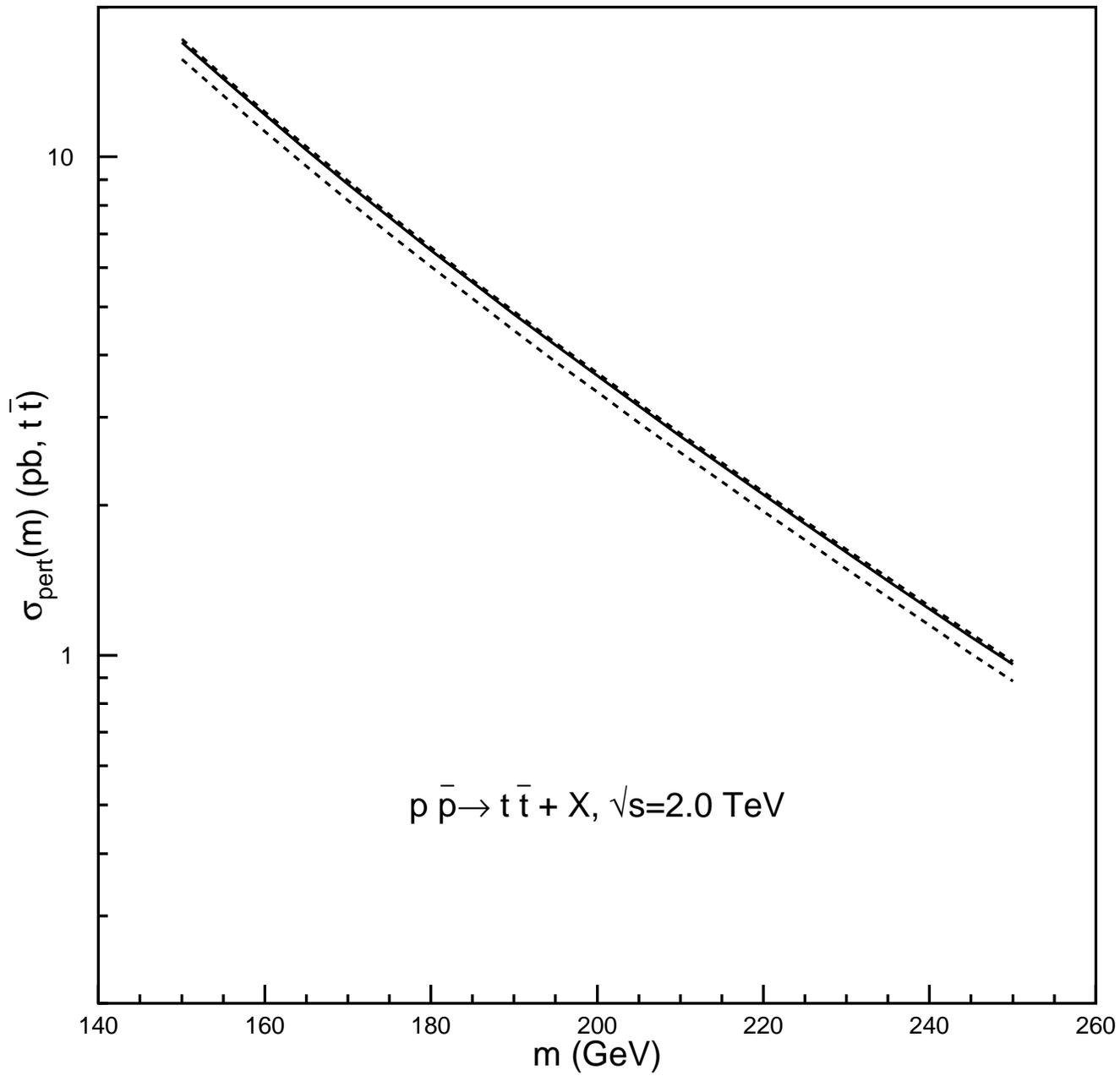
\begin{figure}
\caption{Inclusive cross section for heavy quark production at 
$\protect\sqrt{S}=2$ TeV in the $\overline{\rm MS}$ scheme. The dashed curves 
show our perturbative uncertainty band, while the solid curve is our central 
prediction.}
\label{fig6}
\end{figure}



\begin{references}
\bibitem{ref:mowidth}
Since the width of the top quark provides a physical definition of the 
perturbative boundary, it would not seem justified to integrate into the 
part of phase space that is above the value of $z$ ($z_{max}$) and below the 
value of $\eta$ that correspond to this natural limit, as is done in 
Ref.\cite{ref:catani}.
\bibitem{ref:deriv}
The essence of this derivation may be found in Eqs.(122), (129), and (130) of 
Berger and Contopanagos, Ref.~\cite{ref:lgpaper}.
\end{references}
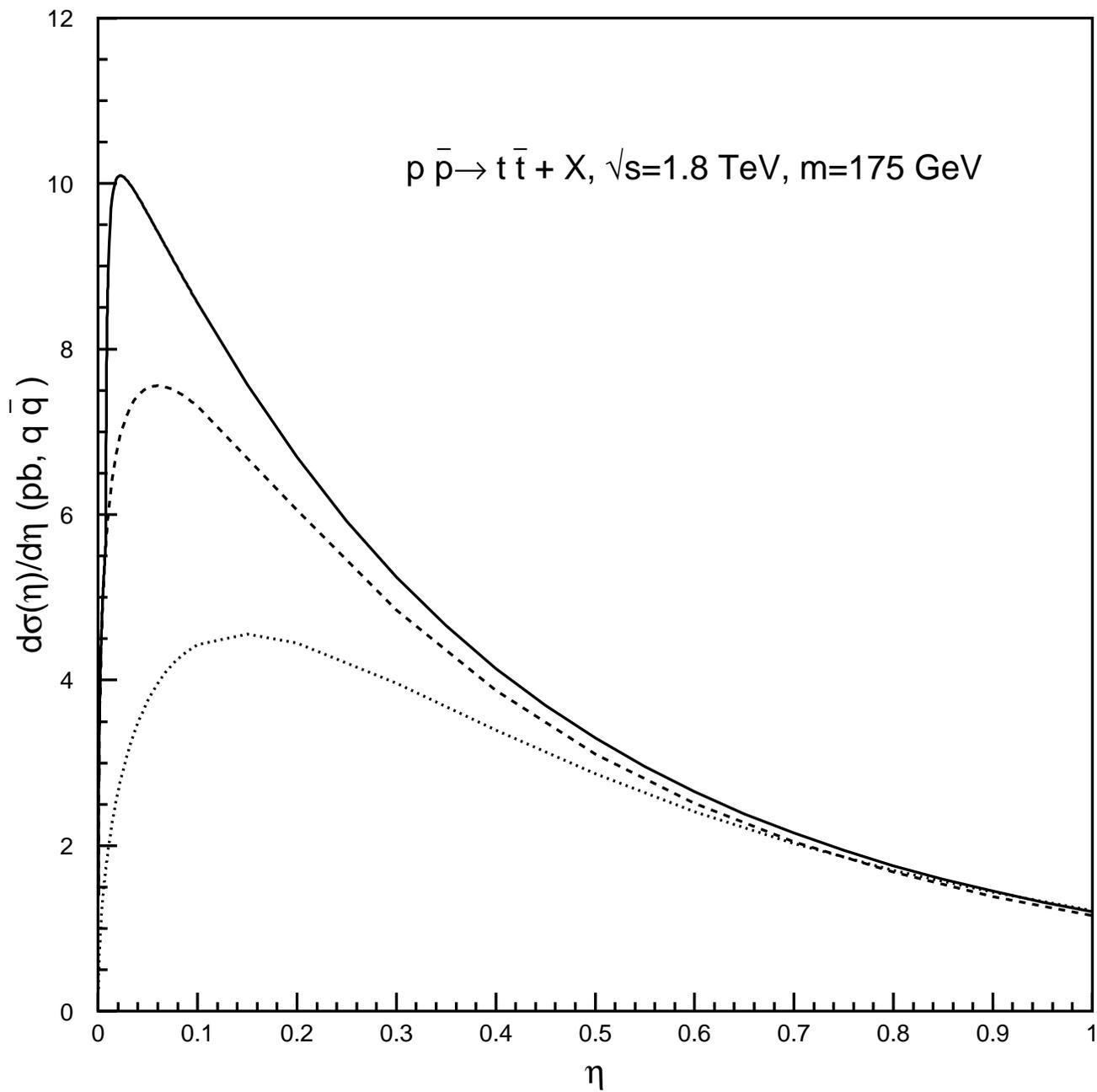
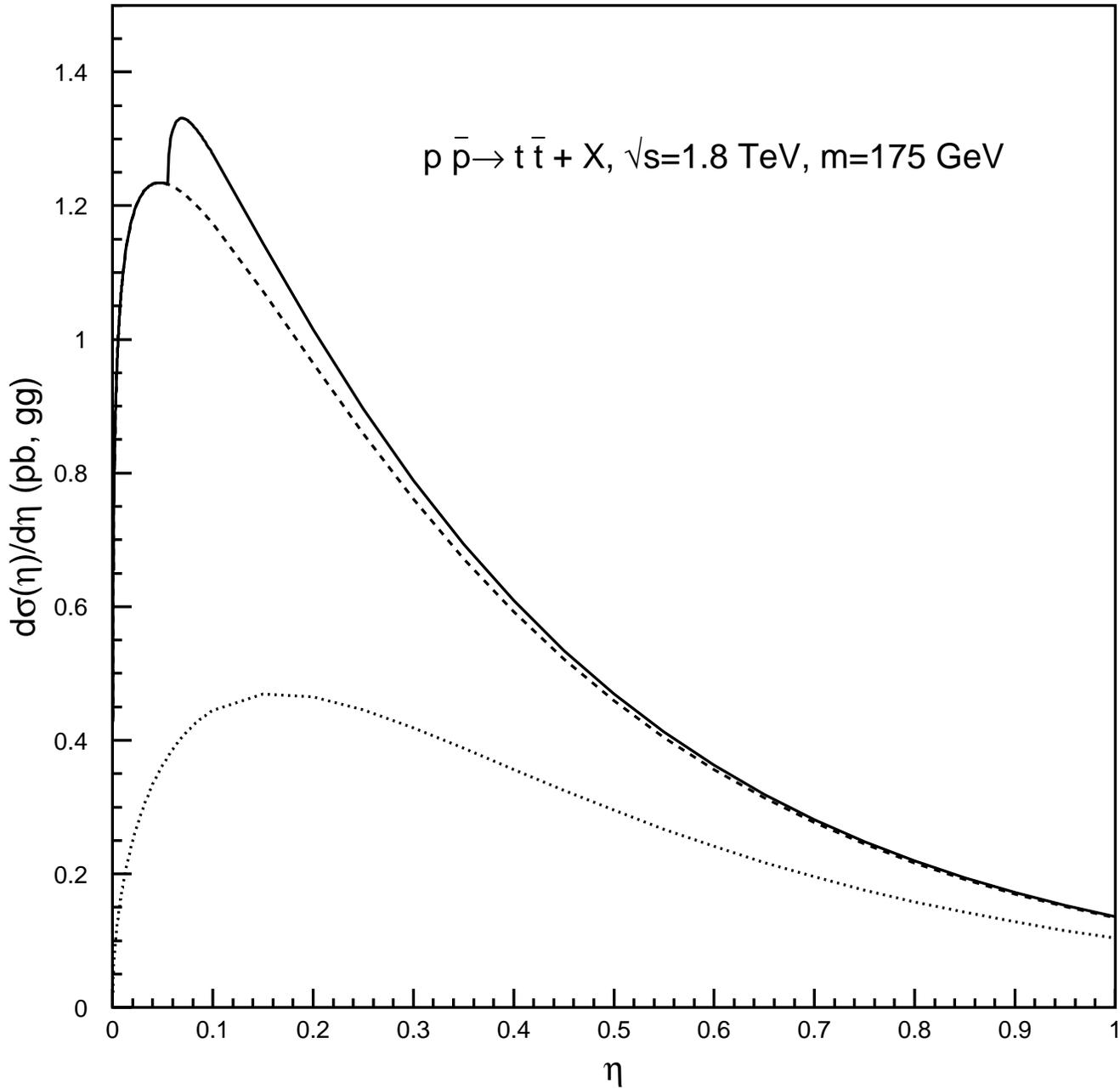
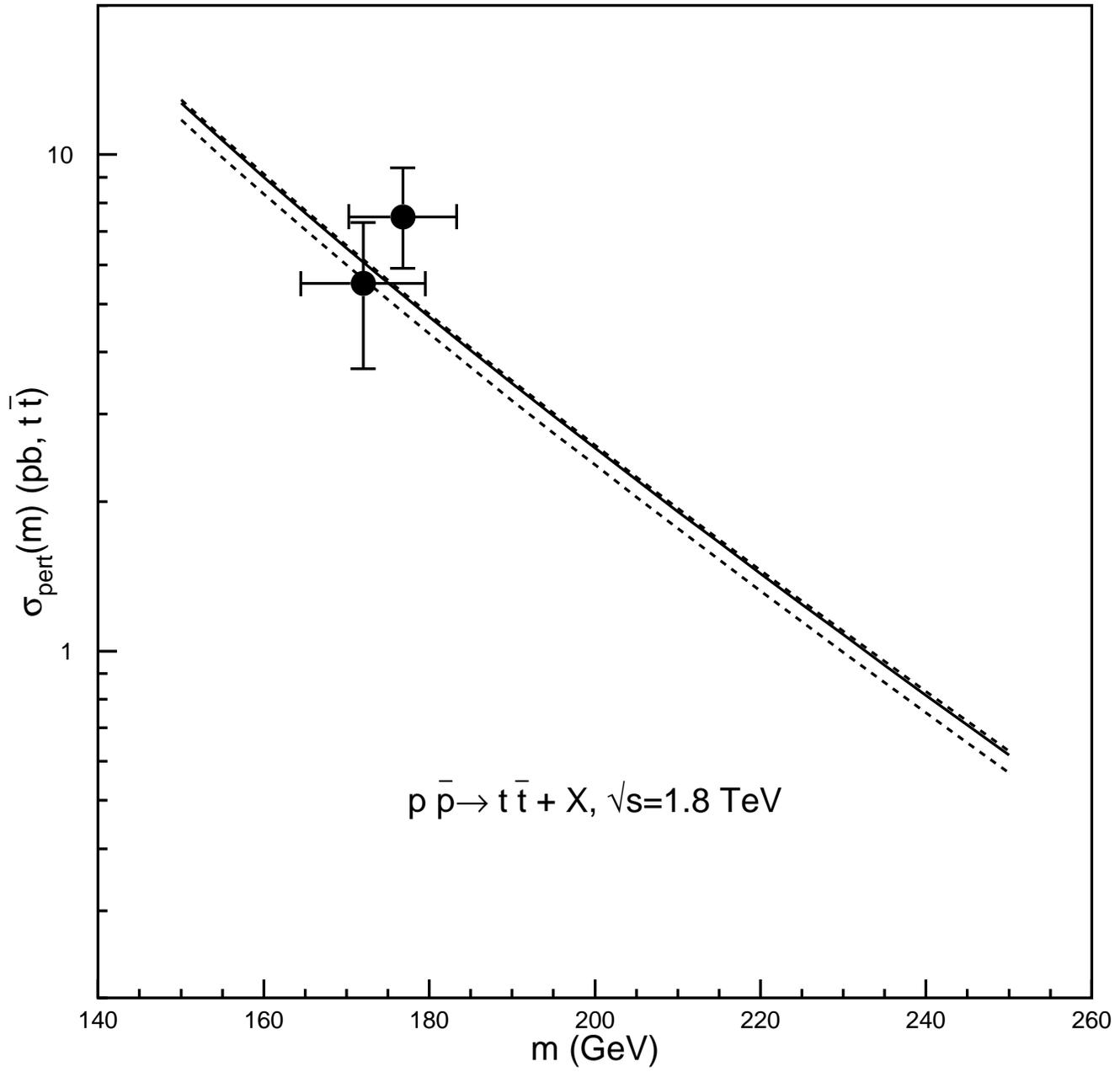
\end{document}